**Models and numbers: Representing the world or imposing order?**

Matthias Kaiser*, Tatjana Buklijas**, Peter Gluckman**

*Centre for the Study of the Sciences and Humanities, University of Bergen

** Koi Tū Centre for Informed Futures, The University of Auckland

**Corresponding author:**

Professor Matthias Kaiser

Centre for the Study of the Sciences and Humanities (SVT),

University of Bergen

Email: matthias.kaiser@uib.no

Phone: +47 917 33 928

**Acknowledgments:**

MK gratefully acknowledges funding from the Norman Barry Foundation which enabled his stay at the Koi Tū: The Centre for Informed Futures, Auckland, and his contribution to this paper.

**Conflict of interest:**

No such conflicts.

**Abstract**

We argue for a foundational epistemic claim and a hypothesis about the production and uses of mathematical epidemiological models, exploring the consequences for our political and socio-economic lives. First, in order to make the best use of scientific models, we need to understand why models are not truly representational of our world, but are already pitched towards various uses. Second, we need to understand the implicit power relations in numbers and models in public policy, and, thus, the implications for good governance if numbers and models are used as the exclusive drivers of decision making.

**Introduction**

This essay was inspired by the profusion of epidemiological models appearing in the public space since the onset of COVID-19. Most of them claimed to be aimed at predicting the mortality of the disease as it spreads "naturally" or against a variety of interventions that societies and governments could impose (Ferguson 2020; University of Washington Institute for Health Metrics and Evaluation 2020; Lourenço 2020; James 2020; Binny 2020; Public Health Agency of Sweden 2020). When facing a disease so new as COVID-19, the scenarios that these models could provide were highly influential. At the same time, predictions arising from different models diverged so widely and spurred such intense discussion that a public debate emerged, in media and in academic journals, around the benefits and costs of the use of models (McCoy 2020; Daalder 2020; Kottasová 2020; Sayburn 2020; Rhodes, Lancaster and Rosengarten 2020). Such diversity and contestation in the public space could allow political decision makers to ignore or dismiss the science and contribute to undermining trust in scientific insights.

Constructing mathematical models to predict the spread of an infectious disease within a population is not a new practice (Ross 1911; Kermack and McKendrick 1927). Mathematical techniques to explain and predict disease trends were part and parcel of a biological, evolutionary, and ecological perspective on the infectious agent which emerged in the early twentieth century to contrast the older medical approach focused on disease control (Mendelsohn 1998; Honingsbaum 2020). Although mathematical models have been used for decades towards a variety of purposes,[1] tensions between mathematically- and medically-trained epidemiologists, regarding what constitutes more robust knowledge and which approach has greater objectivity and value, persist (Amsterdamska 2005).

Controversy around the decisions made by policymakers made based on modelers' predictions is also not new. In the US in 1976, prediction and extrapolating assumptions of a swine influenza epidemic and mortality based on the 1918 epidemic led to an unprecedented mass immunization effort (Dehner 2012 pp. 2–6).[2] But the killer never came, and the vaccine had a serious side-effect: a neurological condition called Guillain-Barré syndrome. An investigation into decision-making that resulted in these outcomes stated, devastatingly, that the swine flu program was characterized by "overconfidence by specialists in theories spun from meagre evidence" and "failure to address uncertainties in such a way as to prepare for reconsideration" (Neustadt 1978).[3] In the early 2000s, a major controversy emerged around the unusually large outbreak of the foot-and-mouth epidemic affecting British agriculture. This epidemic zoonosis came on the heels of the major crisis caused by bovine encephalopathy (BSE) and the associated human Creutzfeld-Jacob disease in the UK. Taking the leadership out of the hands of the Ministry of Agriculture, Fisheries and Food and into the newly established Science Advisory Group for Emergencies (SAGE) was supposed to place decision-making on a more "scientific" footing. The rapidly developed models were the basis of the decision to undertake widespread preemptive culling of many animals. Post-foot-and-mouth, modelers got a key role in SAGE; indeed, Neil Ferguson headed the modelling in both foot-and-mouth epidemic and early stages of COVID-19 in the United Kingdom (Millstone 2001; Ford 2020). This continuity between the two big crises could lead us to think that using mathematical modelling in foot-and-mouth was an undisputed success. Yet early on models were criticized for the poor quality of data and modelers seen as lacking the first-hand experience of the disease and its social impact (Campbell & Lee 2003; Kitching et al 2006; Mansley et al 2011). Indeed, many of the features of the foot-and-mouth crisis—contestations around expertise, or the way that media first built up and then destroyed models and modelling—re-emerged in COVID-19 (Bickerstaff 2004, Nerlich 2007).[4]

---

[1] For example: to understand and compare the disease spread; to compare the effects of prevention and control procedures such as screening (testing), tracing the infected and their contacts, and immunization; and to refine the prevention strategies e.g. to find the optimal time of immunization (Hethcote 1989).
[2] The prediction was based on a very simple calculation: in 1918 the death rate was ca 500k, but since the population had doubled, and thus one might expect a death rate of 1 million.
[3] However Silverstein (1981) was less critical of the role of experts.

Yet while modelling infectious diseases and indeed epidemics is not new, modelling **pandemic** outbreaks of human infectious diseases remains a relatively novel undertaking. Not available in the influenza pandemics of 1918, 1957 or 1968, the first time that it was applied to guide international emergency responses was in the early 2000s. Thincreased interest in influenza virus over the past decade, around the outbreaks of highly infectious H5N1 (avian) influenza converged with the availability of advanced computational technology and methods (Hatchett 2007). They were first applied in the H1N1 pandemic of 2009 (Lee 2013). These models were deemed useful for highlighting gaps in data, characterizing the role of the so-called "non-pharmaceutical interventions" (measures such as social distancing, shown to be so important in the 2020 pandemic), prioritizing populations for vaccination, and also as a heuristic for decision-makers. Yet some of the objections raised around modelling in 2009 emerged again in 2020: the mismatch between expectations and interpretations from the policy side versus the uncertainties and implicit assumptions on the side of model developers.

---

[4] On re-emergence of the debate in 2020 see Adams (2020).

While in 2009 epidemiological models were "not generated, shared or disseminated in time" (Lee 2013 p. 1014), in the COVID-19 pandemic the sheer speed of production, variety, public impact and discussion of mathematical models has been astounding. Many of these models were directly released to the public, few were subject to formal peer review before release and many did not have the detail necessary to evaluate the assumptions involved. Some, at least, influenced policy, directly or indirectly, depending on their source and whether they were centrally requested or simply proffered in the hope of influencing public opinion and, ultimately, policy. Worries that they may have done so to the exclusion of other kinds of information and detriment of public decisions have been voiced (Sridhar 2020).

Graphical representations of models were shared through regular and social media. Sometimes taken out of the context and divorced from the detail on model limitations and aims, their message was simplified and sometimes changed. At the same time, whether these models did indeed influence political leaders' choice of the epidemiological strategy (as they decided between "herd immunity", versions of suppression such as "flattening the curve" and elimination) is not well understood .[5]

To contextualize and discuss the ensuing problems, this article consists of three main sections. The first (epistemology/philosophy of science) is an overview of the mathematical models in science – their purpose, use, and epistemological limitations, ending with a short examination of the epidemiological (mathematical) models. In the second part we place mathematical models in the contemporary media ecology and discuss them in relation to the new publishing practices in an online era and with regard to the rhetorical power of the number. In the third part we propose recommendations that scientists and communicators should try to follow in making and communicating models.

**On objectivity, data, systems and models**

To understand science/scientific information, one needs to break through the shielding rhetoric of "objectivity" – a culture of relating to the world, which in the nineteenth century was meant to replace the dominance of subjectivity (Daston 2007). Although the goal of the culture of objectivity has always been to remove the subjective human as much as possible – for instance through the mechanical means of representation, measurement and calculation; by using cameras, digital calculators and computers instead of the eye, hand and brain – complete objectivity has proven an illusion (Rose 1983; Daston 2007). Scientific problems are always pursued with a degree of perspectivity (Giere 2010; Saltelli 2020a). We "measure" an earthquake, but we know that our measurement of its intensity is built upon various assumptions and abstractions (Chang 2004).

---

[5] In December 2020 the Prime Minister of New Zealand explained how the epidemiological models combined with the knowledge of national healthcare capacities led to the switch from the widely publicized "flattening the curve" to elimination strategy in March/April 2020 (Perry 2020). To our knowledge other international political leaders have not offered comparable insights into their strategy choices. On epidemiological strategies used to tackle covid-19 internationally, see Allen et al (2020).

Hence even data are in the final instance partly constructions, and not mere observed facts of nature (Fleck [1935]1979; Latour 1979; Kaiser 1991). That does not imply total malleability, nor a necessary multiplicity of viewpoints. In science there is a widespread intersubjective consensus on virtually all data; science "functions" not because it is objective, but because its data, assumptions and abstractions are intersubjectively shared among its practitioners. Data are reflective of reality but that reflection depends on multiple assumptions and methodological considerations. Data appear to us in ways we are interested – and which we sometimes can measure - in order to understand a certain segment of reality. They are the epistemic priors upon which (virtually) all our theoretical understandings in science are built and justified.

We use data to represent the phenomena we study. **Phenomena** are the relatively stable or fixed features of reality that typically are believed to display causal mechanisms, and are themselves believed to be causally related to other phenomena (Kaiser 1995, 1996). Our scientific theories are about phenomena. In this way we construct systems as representations of the world as abstractions from the data with all the limitations discussed above. A **system** is simply a collection of entities which interact with each other by a number of relationships (generally simplified from the many relationships that might exist), and which includes one or several phenomena (von Bertalanffy 1968 pp. 30–53.). Even when systems are in reality 'open', they tend to be described as 'closed', where all the interactions outside the system are excluded. A system can be expressed in various ways, as a mathematical structure, as a graphic figure, or as a mechanical machine.

An example of a system can be the set of tectonic plates and their movement on the seafloor, comprising, among others, the phenomena of drift, earthquakes, volcano eruptions, transform faults, magnetic reversals, and pole wandering. A simpler example of a system could be mobility and migration among people, consisting of a stratified set of people, places, and crisis events like hunger, war, or political unrest. The crisis events are the phenomena that trigger causal mechanisms in the system. All systems are in a very basic sense already a **model** of reality. Any such system is taken to represent certain features of reality by way of a **similarity relationship**. They resemble what they are supposed to depict in salient ways, but never in all respects, and they initially stay close to the ensemble of the available data. They are like maps in relation to a landscape. They are constructed with a certain viewpoint in mind, given our observations and given that simplification ignores potential minor interactions that might indeed matter. They generally assume a closed rather than open construct. Sometimes it may be hard even to detect how data are included in these constructions of a system.

However, when scientists talk about modelling, they typically go a step further. A model is a mediator between phenomena (reality) and theory (abstraction) (Chadarevian 2004; Gelfert 2016). Historically many models were physical, three-dimensional objects, even living cells and animals, used for different purposes (Chadarevian 2004). One important use is testing: engineers use down-scaled physical ship models to test their properties in turbulent waters. Medical scientists use animal models, i.e. real animals representing certain human properties, to test toxicity or other medical conditions. Organoids or three-dimensional stem-cell grown cell-structures representing human organs are used to test treatment of diseases. But there are numerous other relevant purposes where models can help. Wooden, wax, or computer models of the body historically solved the problem of the difficult and problematic access to human embryos and bodies while also providing pedagogical tools more vivid than the flat images (Maerker 2011). Models can guide field studies and help develop theory. Sometimes their role as predictive tool is stressed (Levins 1966).

A **scientific model** nowadays is usually a mathematical model. It is constructed on the basis of a given data set. The data set describes one or several phenomena. These are then the main categories of a system, which we want to explore further in specific respects. A model depicts the relationships in the system as dynamic or static functions, differential equations, or complex functions, such that one can move forward or backward (in time and/or space) in relation to the given observations. We want to understand the dynamics of a system and therefore we construct a scientific model. Some state this representation function as an isomorphism between model and target system (Bailer-Jones. 2003), but below we shall follow others who see this as an incomplete characterisation of models (Giere 2004, 2010; Knuuttila 2005). Ultimately "a model is an abstract representation of a system or a process" (Turner 2015).

There is an epistemic functionality between the input and output of a model: we need comprehensive and reliable input (data and parameters of the system) in order to have a chance to generate reliable output. However, even with good input we do not always get the good or intended output, since we can only construct models over segments of reality (systems).

A general rule in model construction is to start with an initial reflection on the framework of the model:

(i)     Clearly define the purpose and the scope of the model,

(ii)    Define the inherent limits of the available information and measurements that go into the model construction, and

(iii)   Consider the possible consequences of model error on the output and its usability.

This initial consideration provides guidance and will help specify the system's boundaries as well as the model components, i.e. the state variables, and their relationship to each other. In mathematical models, it will clarify whether we are dealing with an analytical or numerical model. Differential equations in numerical models can help representing non-linear complex systems by providing state approximations.

It is essential to realize, then, that models are not true or false (Oreskes et al. 1994), not even in the sense of being closer to the real system of the world studied. They are only better or worse in relation to the purpose they are designed for. This is also in line with the post-normal science framework (Funtowicz 1993) which stressed quality, perceived as fit-for-purpose, as the arbiter of good science. Thus, the characteristic model definition rests on a pragmatic relation, put forward by Ronald Giere (2004, 2010), here adapted from Knuuttila:

> ***A scientist F uses a model M to represent features of system S for the purpose of P*** *(*Knuuttila 2005)**.**

This multi-functionality of models is in one sense an advantage of models. At the same time, their versatility is also a serious restriction in the sense that the choice of function determines the qualities of the model.

There are interesting and consequential trade-offs to be made in model construction. Levins (1966) observed a principal trade-off between *generality*, *precision*, and *realism*. One may sacrifice generality for the sake of realism and apparent precision, and thus calibrate the model so that it closely resembles all the data we have of the target system.[6] But doing so will make it less applicable for other localized systems. Or, alternatively, one may aim at generality and precision but sacrifice realism. Here we may be content with a good average result fitting a larger family of systems, but no single one fitting precisely. Finally, one may sacrifice predictive power to realism and generality. This is often the case when qualitative results are good enough for the given purpose. "Many theoretical models fall into this category, and they make very general predictions that are not directly applicable to a particular place or set of measurements". See also **Figure 1**.

---

[6] We follow Turner and Gardner (2015) in this presentation of the trade-offs.

[INSERT FIGURE 1 HERE]

There are, other tradeoffs as well. Complex models are seen as more accurate and simple ones as more general. A lack of detail is seen as causing systematic bias in predictions—but adding detail to a model does not guarantee an increase in reliability unless the added processes are essential, well understood, and reliably estimated (Turner 2015, p. 70). O'Neill's conjecture was that there may be an optimal balance between model complexity and model error (**Figure 2**).

[INSERT FIGURE 2 HERE]

Our systematic point then is that model utility is crucially dependent on the uncertainty trade-offs in their construction. Uncertainties are built up from data collection, onwards via scale, scope and boundaries of target systems, to parameter selection and design functionality of the model. Communicating these trade-offs and uncertainties is a moral imperative when handing models to potential users of them. This is especially true when these models are to be used for predictive purposes; with highly complex target systems, or systems with largely unchartered causal interactions, the predictive value rapidly decreases even if the model accurately depicts current processes.

**Epidemiological models used during a pandemic**

Epidemiological models are mathematical models: systems of differential equations used to describe how disease spreads in a population. In their simplest form they describe the spread of an infectious agent through direct contact by dividing the population into separate classes which change with time: **Susceptible** class (those who can be infected but aren't yet); **Infective** class (those transmitting the disease to others), and **Removed** class – those who died or acquired immunity. These SIR models purport to predict the rate at which the susceptible class moves into the infectious, and then the infectious into the removed. And while these rates depend on the biological properties of both the infectious agent, the process, and the host (the properties of the microbe; mode of transmission; length of incubation/latent period; susceptibility and resistance), they are also a function of a range of behavioral, social, cultural and material/technological factors.

In epidemiological models, some of the above variability is reflected in differentiating between best and worst scenarios (and various other cases in between). One can do this by e.g. changing the statistically mean number of infections from a given source ("basic reproduction number") $R_0$ e.g. from 1.4 to 2.8 people infected[7], which then represents a global average for a population (Holmes & Masuda 2015). But calculating this rate is beset with some methodological difficulties and implicit abstractions or idealizations: "one needs to assume that every pair of individuals has the same chance of interacting at any given time. In fact, interaction rates depend on pairs of individuals—people living in the same city are more likely to interact than those living in different cities" (Holmes 2015, p. 2).

---

[7] This was the estimated range of $R_0$ for the 1918 swine flu epidemic (Coburn 2009).

Typically, the calculation of $R_0$ assumes a totally susceptible and openly interacting population which may not be true, based also on behavior patterns and variance in individual immune responses.

Furthermore, the probability of an infection given some appropriate contact between individuals is highly uncertain, given that this implies knowledge about the previous number of potentially contagious events between infected and non-infected people, and where the number of infected people without disease symptoms usually is unknown. There are also interdependencies to be aware of. If $R_0$ <1, then the total number of infections $\Omega$ is independent of the population size N, while when $R_0$ >1 then $\Omega$ is proportional to N, i.e. only a finite fraction of N can be infected. In principle, there are both spatial and temporal variations in contagious events which introduce basic uncertainties in the calculation of $R_0$. This uncertainty about the correct $R_0$ rate is then usually communicated by modelers as variation in the predicted outcomes, and with stochastic simulations one may partially compensate for the lack of knowledge of basic parameters of agent or contagion behavior. Yet, in order to know what trust to put in these predictions, we would need to know the scope and scale of the model, the data that went into it, and the assumptions built into it, for instance about the transmission types.

The calculation of the basic reproduction number is the basis for calculating $R_{eff}$, i.e. the rate of effective infections given a number of countermeasures to contain infections but also variations and constraints in people's reactions to containment measures, elementary unknowns of the prevalence of the virus in the population and the reliability of various testing kits, the indicator of stringency of measures,

under- or over-reporting of health care units due to external factors. How many super-spreaders do we work into the system, and what is their effect? With a different $R_{eff}$ and an identical $R_0$ one may hope to see how one country's counter-measures are more effective than another country's interventions. With good data this could be a good use of an epidemiological model. But it does not follow that even a good $R_{eff}$ is a reliable predictor of future infections. We can see the function of scope and scale of data as the crucial restricting element of predictive utility, together with the in practice unrealistic assumption that the targeted system of $R_{eff}$ is a closed system – in reality, it is always open.

We need to understand the gross abstraction from reality that every model needs to build upon. Even a more elaborated agent-based model is not populated by real people (Snijders 2010); it is populated by types and network characteristics, as friendships and collaborations, risk-takers and risk-avoiders, and / or rule-compliers and rule-breakers, or infected, non-infected, immunized, and in treatment.[8] To gain more realism, one would also have to add spatial and temporal differentiations, while all these further specifications lead to a culmination of uncertainties, given that accurate measurements of all of them is unrealistic. Again, we see the trade-off between realism and generality.

There are many unknowns in such model building, and in this sense, they usually cannot adequately represent the state of knowledge. Indeed, one of the major uses of models is to point out where new data is needed. They can be used to suggest how the disease *might* evolve. They can also force explicit statement of assumptions about the biological and socioeconomic mechanisms which influence disease spread. Yet what happens in the times of a crisis is that various possible uses are mixed and under-communicated. The inbuilt uncertainties disappear and sensitivity analyses are rarely included in times of crisis (Saltelli et al 2008).

---

[8] Cf. also the interesting COVD-19 typology suggested recently (Lam 2021).

In conclusion, models always represent a high degree of inbuilt uncertainty, and some models, as e.g. models with non-linearity, exhibit huge inherent uncertainties, relevant in many contexts. All of this constrains significantly model reliability as predictive and even as explanatory tool. Typically, we will be confronted with a diverse set of models, built on different assumptions, and predicting different outcomes, and without the enabling information to make critical use of them. When data sets are poor, and knowledge of the causal dynamics of the target system is highly restricted and hypothetical, then complex mathematical models rarely can help us out of the conundrum of urgent complex decision making.

**Epidemiological models and the novel practices and tools of science communication**

In the COVID-19 pandemic, mathematical models seemed to appear with breath-taking speed and were fed to the policy maker and general public. There are assumedly multiple explanations for this overload of model outputs. Some of them may be sociological, political and also psychological: competing between scientists for access to power, securing research funding for the future, improving the impact accounting now in operation in much science policy; and sometimes requests by governments. Clearly, policy makers needed to have a sense of the likely path of the pandemic though it might have sufficed to heed the warnings of the People's Republic of China's experience, showing that the approximate $R_0$ in a naïve population would overburden healthcare systems.[9] While full papers may have explained the range of possible scenarios and assumptions, media and politicians often appeared not to adequately understand this aspect of modelling.[10] These models were used by both scientists and politicians to argue for how many lives have or have not been saved as if they described reality and in the absence of any counterfactual being possible. With all the uncertainty removed in the process, the expected (normal) divergences in models' predictions now appear as contradictory signals from a community that the public and the decision-makers perceive as scientific experts and from whom they seek guidance in the midst of the crisis.

In the rush to communicate models quickly, models are published through public releases, blog or preprints rather than going through the customary procedure in academic science: submission to a journal, followed by at least one round of peer review, and then publication. The critical multi- or transdisciplinary vetting and quality assurance of the model as a constructive instrument in science advice to policy and government is omitted (Gluckman 2020).

---

[9] Paying close attention to the situation in PRC seems to have formed the basis of Taiwan's successful response (Lin 2020).
[10] So Ferguson et al (2020) or "Imperial study" was often understood to predict 510,000 deaths from covid-19 unless the government abandoned its current ("herd immunity") strategy. See e.g. Bostock (2020).

Peer review is often seen as the cornerstone of academic science that has been around since the first professional journals, such as *Philosophical Transactions* of the Royal Society. Yet historians have shown that the uptake of peer review was patchy until relatively recently (Csiszar 2016). In the 1970s and facing a prospect of diminished public funding as well as increased oversight by politicians and policymakers, American scientists cast "peer review" as the crucial process that ensured the credibility of science while at the same time keeping the control in their own hands (Baldwin 2018). The dominance of American science and American scientific journals then helped this model to spread around the world.

Yet in recent decades, new modes of direct publication via open-access depositories of preprints such as arXiv, bioRxiv and medarXiv, have sprung up. Manuscripts placed into these archives become permanent part of the scientific record and citable with unique DOI before going through the peer review. These depositories laud their publication model as enabling authors "accelerate the speed at which science moves forward" (https://www.thelancet.com/preprints). The expectation is, however, that the preprints are just temporary substitution for the fully reviewed and accordingly revised final version. Indeed publishing preprints is meant to informally expand the pool of peer-reviewers with the goal of a well-validated final version.

In the pandemic, however, the difference between preprints and final versions is blurred. Yet, omission of peer review can have serious consequences. A study that compared the mean and range estimates of basic reproduction rate of studies published between 23 and 31 January 2020 as preprints (without peer review) with those that did go through peer review found the mean $R_0$ of peer-reviewed studies lower, and the range of estimated $R_0$ narrower, than that of the preprints (Majumder 2020). The same study noted that it is the preprints that are "driving discourse related to the ongoing COVID-19 outbreak," and that it is "likely that preprints are also influencing policy making decisions." (Majumder 2020, p. e628). With non-peer reviewed models, which by definition exist a step away from "reality", the danger of pushing the course of intervention in the wrong direction is further heightened.

Finally, publishing preprints is just part of the vast new ecosystem of digital communication (Davies 2016. pp. 5–6). Social media, especially Twitter, offers a communicative mode that is even more rapid and direct than the online repositories (Puschmann 2014). But distilling a model with all of its assumptions and limitations into 240 characters is challenging. Models, then, tend to be reduced to the most dramatic prediction (the worst-case scenario) and the most vivid graph; as the next section will show, it is difficult to overestimate the rhetorical power of numbers.

**Rhetorical power of numbers**

It is crucial to realize the rhetorical power of communicating numbers generated by models to decision makers and the general public. Particularly in a crisis, decision makers directly base decisions and steps to contain the crisis on available numbers. The dynamics of $R_0$ is typically key ("flattening the curve"). Therefore, the temptation is big to extend the use of numbers generated from data and models even more deeply into areas where the uncertainties are overwhelming. It takes courage to withstand such a call to put numbers on the table when they amount to pure guesswork. According to a TV clip on the exchange between Rep. Glenn Grothman and Anthony Fauci, the American infectious disease expert and lead member of the White House Coronavirus Task Force, Fauci when pressured by the politician Grothman in congress on the number of deaths expected for COVID-19 courageously stated "*There is no 'number-answer' to your question*". His response resembles the answer that in 1976 James Dickson gave to David Mathews (Secretary of HEW in the Ford administration) inquiring about the probability of deaths in the expected swine flu epidemic: "*Unknown!*" There is an inherent ambiguity and inherent power dynamics to this answer. On the one hand, it preserves scientific integrity in not passing mere guesswork as scientific information. Yet on the other hand, it hands to politics a clear necessity to prepare for the worst imaginable case, else the political price will be too high. This was also Gerald Ford's conclusion about the necessity of preventive vaccination in the 1976 swine-flu episode (which never came): "I *think you ought to gamble on the side of caution. I would always rather be ahead of the curve than behind it"* (Neustadt 1978, p. 22)*.*

This then introduces a conundrum in the process of science advice to politics: *The most truthful answer may not be the most useful answer!* It also illustrates the difficulties in practiced co-responsibility. In not giving the number estimate that the politician would like, one implicitly may leave it to the politician to prepare for the worst case; nobody wants to be blamed for not taking all the necessary precautionary steps in spite of being forewarned of a possible pandemic. Understanding the political dynamics should be a factor considered by scientists in their risk communication. The way to account for this is not to put a number forward that is unwarranted, but to prepare the stage for an adaptive and proportional risk management. This communication then has to be clear on the vast uncertainties and the values at stake (Funtowicz 1990, 1993), and pave the way for adaptive strategies.

Furthermore: some numbers are available, e.g. number of deaths in other countries. One does not need sophisticated models to see the serious spread of the infection in the curves that depict infection rates and deaths. Many politicians in the West commented that the Italian situation spurred them to action, while their peers in South East Asia, SARS fresh in their mind, reacted even earlier to reported numbers. In the COVID-19 case, the state of Taiwan, Vietnam and South Korea with their early and comprehensive testing had much better data on the infection rate than e.g. Italy. What this shows is that numbers indeed are a powerful trigger for politics, and in times of an imminent pandemic reporting by numbers may be one of the most effective means of science and risk communication, alongside visual means such as dramatic photo and film material.

The accumulation of these data as a curve in a (logarithmic) graph provides relatively trustworthy and robust information on our current – and usually past – situation. While this cannot uncritically be taken as established facts, this is as good as it gets when it comes to quantified information. And this is normally all the information that a decision maker really needs. In other words, just using ordinary graphical depiction of data, and some household statistics can give us numbers that are relatively robust, and mostly sufficient for initial decision making: a version of Herbert Simon's term "satisficing". In effect, we argue that in times of an imminent

Page 22

crisis or in its earliest stages, simple heuristics may be more effective than sophisticated formalized modeling (Todd 2000).

As a pandemic progresses, the kind of information important for a decision maker typically changes. It moves from (i) information on the growth and geographical spread of infections, to (ii) information about the time it takes for the doubling of infections, to (iii) information about the best estimates of $R_{eff}$. In addition, (in effect qualitative) information is needed about how effective a certain cluster of measures is to bring the infections down, and therefore which of the measures may be loosened eventually.

There are, however, some basic concerns linked to the use of numbers in politics, particularly the excessive or decisive use of numbers in policy making and as a political argument trumping other qualitative or ethical considerations. The much acclaimed call for evidence-based (or, better, evidence-informed) politics can be suspect in this regard if practiced poorly. Nicolas Rose (1991) in a review article discusses how numbers became an important tool for policy during the 18th and 19th century (cf. also Hacking 1975). They promised to expel subjectivity and arbitrariness out of politics in favor of democratic accountability and reason. Numbers are still seen to promise a certain objectivity in politics and they leave little room for subjective judgement. Rose argues that "democratic power is calculated power, calculating power and requiring citizens who calculate about power" (Rose 1991, p.673). Quantification, numeracy and statistics became instruments of policy making not only for the economy, but basically throughout all areas of state authority, including health. No policy seems well-founded unless it rests on reliable information on the state of the population, which for instance presupposes a census and other forms of collecting data on people. Yet, all such data collection is already inherently biased by values (Douglas 2009). An example is the census question about e.g. ethnicity, rather than race, nationality, ancestry, cast or religion. What is counted is already value-laden, and so are the measures used. In diversified societies the introduction of numbers via statistics represents a means of standardization which makes possible stable and coordinated interactions between widely dispersed sectors (Porter 1996). The complexity of our experiences is compressed into single figures and numbers which makes them appear more objective.

In science, reliable numbers signify a triumph of discovery (the charge of an electron, or the number of genes in human DNA), while in politics numbers should always be accompanied by a red warning lamp: "is it certain, and are there other factors that we need to consider that resist measurement?" The sins of extensive uses of cost-benefit analyses in politics should by now be recognized by most people (Shrader-Frechette 1992). Yet, the rhetorical power of numbers injected in a policy context is not always recognized and contextualized. We certainly do not argue for getting rid of numbers in politics altogether. But one needs to embed these numbers – and of course models – in their appropriate socio-political context of uncertain knowledge and disputed and challenged values.

One problem seems to arise at this point: when we say that these models have to be embedded in the appropriate socio-political context, who is actually doing this, whose responsibility is it in the end? Are we making scientists into shadow politicians and decision-makers? We certainly do not want to imply this. Rather we argue for a double reflexivity. On the one hand, those working on complex quantifications and models may need to control their own enthusiasm, and realize that the complexity, uncertainty and implicit value assumptions in their work needs the filter of peers (typically from many fields) and competent communicators before it is ripe for the space of public discussion of political decision-making (Saltelli 2020b). In other words, they may have very valuable insights to communicate to the scientific community, but these insights need larger support than some selected data and advanced model-fit. Decision-makers need the synthesis consensus, not the most refined numbers which may change as a crisis evolves anyway. On the other hand, our decision-makers need concise evidence-based information, preferably in the form of measurable and understandable units. The specific format may be dependent on the interface of some special bodies, expert groups or science advisors who fill the brokerage function between the science and the politics (Gluckman 2021 in press).

**Concluding viewpoints.**

The uncertainties and different functions of models as discussed earlier, the different layers of uncertainty, and the context of values and socio-cultural factors can help us to design certain ground-rules that should guide model use. We do not claim any normative "objectivity" to the following statements, but we put them forward in the hope that they, as heuristics, may inspire the kind of critical discussion that we surmise is much needed in the science-policy community, and in particular in pandemic management:

- Before numbers and models enter the policy space (particularly in a crisis), and are communicated to the decision makers and the publics, one should make sure that they enter an expert circle with all appropriate areas of expertise at the table for quality assessment. Tunnel views need to be compensated by broad fields of knowledge. These should not only include the range of experts in the domain modelled but also those coming from social sciences and humanities. Choices on what to keep in (what is important in any given context or situation) and what to throw out are political (Leigh Starr 1989; Douglas 2009): thus making these choices explicit is necessary for transparent decision-making.

- Parsimony and simplification should be employed with a view to being just good enough ("satisficing") for the current policy needs; when a simple graph with relatively certain data can do the job, then do not extend this by controversial or uncertain model calculations and spectacular numbers.

- Whenever a model is deemed adequate for policy purposes, one needs to communicate the model's strength, design and weakness; a sensitivity analysis should be routine, inbuilt assumptions and uncertainties need to be explicated, and the degree to which robust data are supporting it.

- While the integration of advanced agent-based behavioural, spatial and temporal parameters may add to the realism of a model, and indirectly to the explanatory power of it, we need to realize that this greater stress of realism comes at the expense of the generality and predictive power of the model.

While in this paper we have argued for a methodological and epistemological examination of models, in particular in relation to their current uses in pandemic politics, our fundamental concern has been with the responsibility and ethics of science when applied into policy processes. We maintain that these ethical dimensions of model use, stemming from fundamental epistemological considerations, are not widely recognized or at least openly addressed in the relevant scientific communities. Guidelines, ethical codes, and best practice models for this field would be very useful.

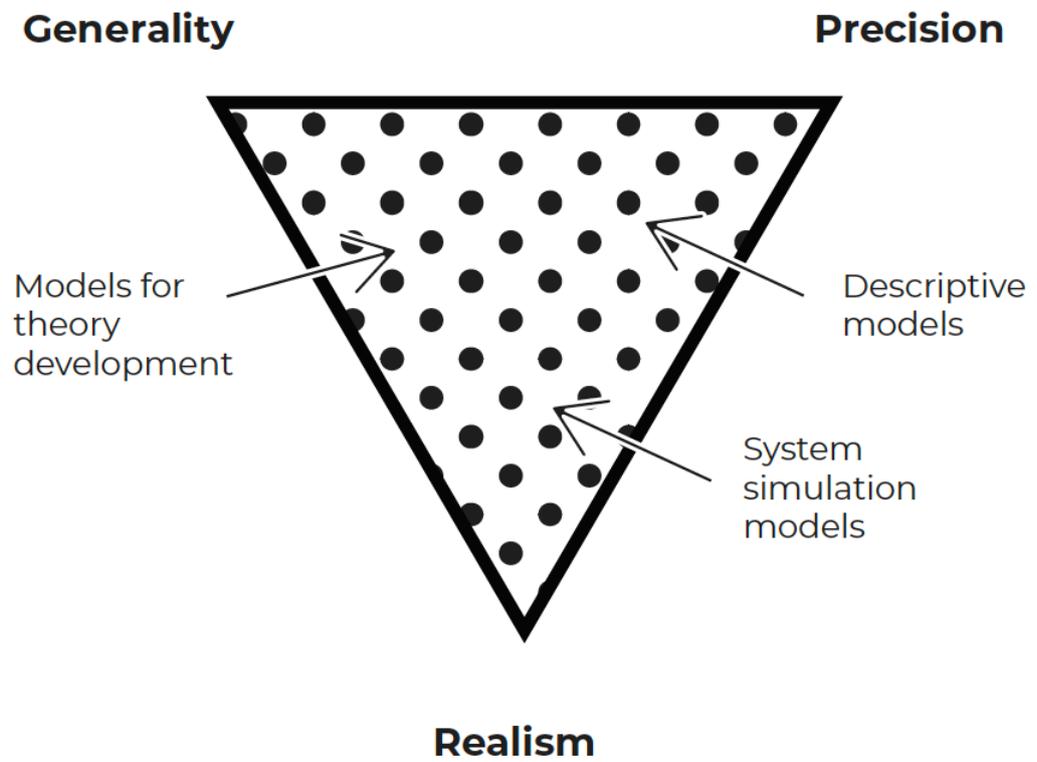

Fig 1: Schematic representation of Levins' (1966) tradeoffs; here drawn after Turner & Gardner 2015, p. 69.

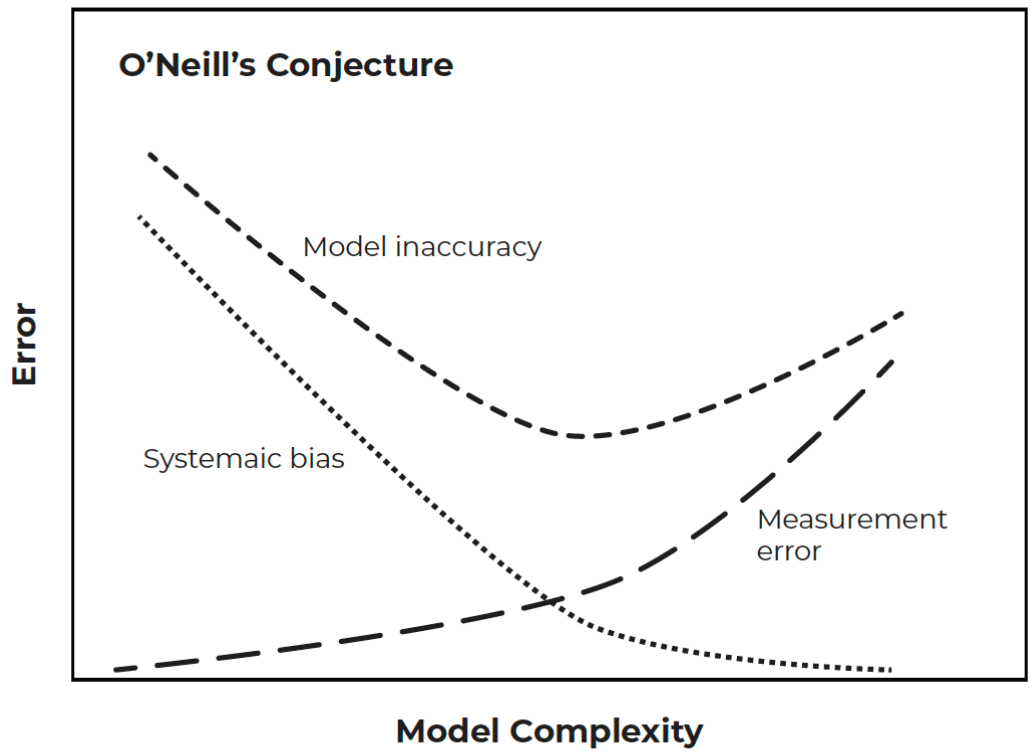

Fig 2. Depiction of the O'Neill conjecture; here drawn after Turner & Gardner (2015), p.70